\documentclass[aps,prl,floats,floatfix,twocolumn,showpacs,superscriptaddress]{revtex4}

\usepackage{epsfig}
\usepackage{latexsym,amsmath}
\usepackage{amssymb}

\begin{document}

\title{Self-similarity of complex networks and hidden metric spaces}

\author{M. \'Angeles Serrano}
\affiliation{Institute of Theoretical Physics, LBS, SB, EPFL, 1015
Lausanne, Switzerland}

\author{Dmitri Krioukov}

\affiliation{Cooperative Association for Internet Data
Analysis (CAIDA), University of California, San Diego (UCSD), 9500
Gilman Drive, La Jolla, CA 92093, USA}

\author{Mari{\'a}n Bogu{\~n}{\'a}}

\affiliation{Departament de F{\'\i}sica Fonamental, Universitat de
  Barcelona, Mart\'{\i} i Franqu\`es 1, 08028 Barcelona, Spain}

\date{\today}

\begin{abstract}

We demonstrate that the self-similarity of some scale-free networks
with respect to a simple degree-thresholding renormalization scheme
finds a natural interpretation in the assumption that network nodes
exist in hidden metric spaces. Clustering, {\it i.e.}, cycles of
length three, plays a crucial role in this framework as a
topological reflection of the triangle inequality in the hidden
geometry. We prove that a class of hidden variable models with
underlying metric spaces are able to accurately reproduce the
self-similarity properties that we measured in the real networks.
Our findings indicate that hidden geometries underlying these real
networks are a plausible explanation for their observed topologies
and, in particular, for their self-similarity with respect to the
degree-based renormalization.
\end{abstract}


\maketitle

Self-similarity and scale invariance are traditionally known as
characteristics of certain geometric objects, such as
fractals~\cite{Mandelbrot:1961}, or of field theories describing
system dynamics near critical points of phase
transitions~\cite{Stanley:1971}. In these cases, objects or physical
systems are intrinsically embedded in metric spaces and distance
scales in these spaces are natural scaling factors. In complex
networks~\cite{review_all},
scale invariance is traditionally restricted to the scale-free property of the distributions of node degrees, which in a vast majority of
complex networks follow power laws of the form $P(k)\sim k^{-\gamma}$,
$\gamma\in[2,3]$. In search for more
complete self-similar descriptions, several recent works~\cite{Makse:2005}
introduced box-covering renormalization procedures, applied them to a few
real networks, and found that certain networks, e.g., the Web and
some biological networks, have finite fractal dimensions and degree
distributions that remain invariant.

Despite this promising progress, self-similarity and scale
invariance of complex networks are still not well defined in a proper
geometric sense. The reason is that many complex networks are not
explicitly embedded in any physical space. As such, they lack any
metric structure, except the one that their graph abstractions
induce by the collection of lengths of shortest paths between
nodes. However, this observable topological metric
is a poor source of length-based scaling factors.
It does not have large lengths as it
exhibits the small-world~\cite{Watts:1998} or even
ultrasmall-world~\cite{CoHa03} property, meaning that the
characteristic path lengths grow not polynomially but
(sub)logarithmically with the network size. The apparent absence of
any other metric structures supports the common belief
that complex networks cannot be invariant under geometric length
scale transformations.

In this paper, we undermine this belief by introducing the concept
of {\em hidden metric spaces} as natural reservoirs of distance
scales with respect to which scale-free networks may be self-similar
at all scales. At the formal level, hidden metric spaces are
variations of hidden
variables~\cite{Boguna:2003b,Caldarelli:2002,Soderberg:2002}.
Specifically, we assume that all network nodes, in addition to
forming an observed network topology, reside in an underlying hidden
metric space, meaning that for all pairs there are defined hidden
distances satisfying the triangle inequality, which can be arbitrarily large. If hidden metric
spaces do exist and play a role in shaping the observed network
topologies, then strong clustering --the high concentration of
triangles-- arises as a natural consequence of the triangle
inequality in the underlying geometry. Therefore, we focus on
clustering as a potential connection between the observed topologies
and hidden geometries.

\begin{figure}[t]
\begin{center}
\epsfig{file=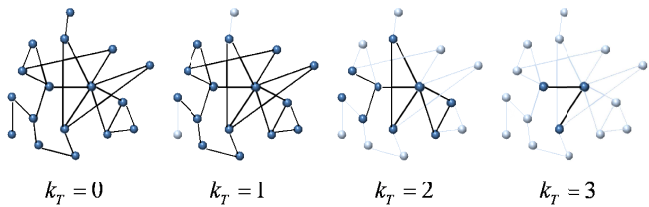,width=8.6cm}
\end{center}
\vspace{-0.5cm}
\caption{Sketch of the degree-thresholding renormalization.}
\label{fig1}
\vspace{-0.5cm}
\end{figure}

Consider the following degree-thresholding renormalization
procedure, which produces a hierarchy of subgraphs within a given
graph $G$ as illustrated in Fig.~\ref{fig1}. For each degree
threshold $k_T=0,1,2\ldots$, first extract from $G$ the subgraph
$G(k_T)$ induced by nodes with degrees $k>k_T$. Second, for each
node in $G(k_T)$, compute its internal degree $k_i$, i.e., the
number of links that connect a given node to other nodes in $G(k_T)$
and, finally, rescale $k_i$'s by the average internal degree
$\langle k_{i}(k_T)\rangle$ in $G(k_T)$ to obtain the rescaled
quantity $k_i/\langle k_{i}(k_T)\rangle$.

We applied this procedure to a few real complex networks and found
that their main topological characteristics --degree distributions,
degree-degree correlations, {\em and clustering}-- are self-similar with
respect to the described procedure: both before and after
renormalization with different values of threshold $k_T$, all these
characteristics closely follow the same master curves describing the
topological structure of the whole subgraph hierarchy.
We next randomized the observed topologies preserving the degree distribution as in~\cite{Bender:1978}, and
found that their degree distributions and degree-degree correlations are still
self-similar, {\em but clustering is not}.

\begin{figure}[t]
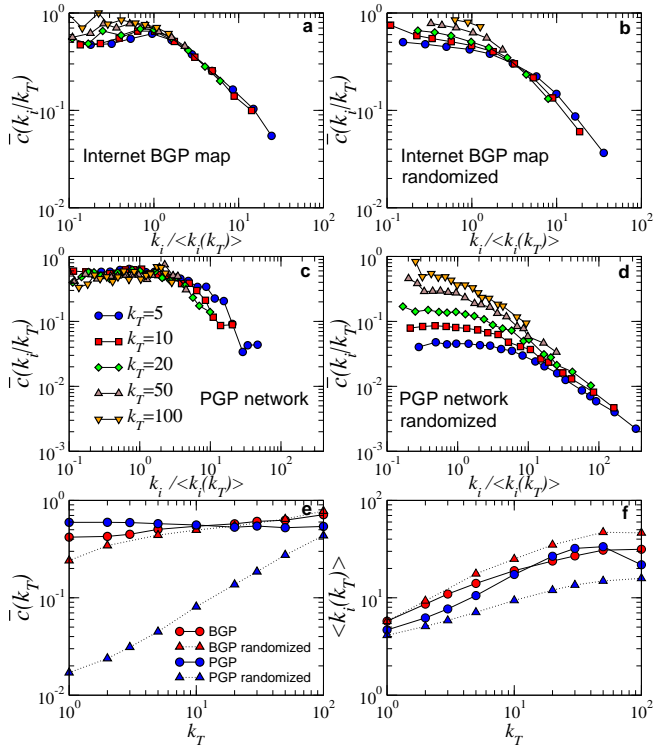

\begin{center}
\begin{tabular}{c}
%\vspace{-1cm}
\epsfig{file=fig2_abcd.eps,width=8.6cm}\\[-0.22cm]
\epsfig{file=fig2_ef.eps,width=8.6cm}
\end{tabular}
\end{center}
\vspace{-0.5cm} \caption{{\bf a-d:} Degree-dependent clustering coefficient as a
function of the rescaled internal degree for the Internet BGP map, the PGP web of trust, and their randomized
versions. {\bf e:} Average clustering coefficient as a function of the threshold degree $k_{T}$ for renormalized real networks and their randomized counterparts. {\bf f:} Internal average degree as a function of $k_{T}$ for the same networks.} \label{fig2}
\vspace{-0.5cm}
\end{figure}

We provide examples of these empirical observations in
Fig.~\ref{fig2}, where we show the degree-dependent clustering
coefficient of the renormalized graphs $G(k_T)$ with different
$k_T$'s for the real and randomized topologies of the Border Gateway
Protocol (BGP) map of the Internet at the Autonomous System
level~\cite{MaKrFo06} and of the Pretty Good Privacy (PGP) social
web of trust~\cite{Boguna:2004b}. Both the BGP and the PGP are
scale-free networks with exponents $\gamma_{BGP}=2.2\pm0.2$ and
$\gamma_{PGP}=2.5\pm0.2$. For brevity, we omit plots showing
self-similarity of degree distributions and degree-degree
correlations. Fig.~\ref{fig2} shows that even though the internal
average degree $\langle k_i(k_T)\rangle$ grows significantly for
all networks, the average clustering
coefficient of $G(k_T)$ as a function of $k_{T}$, $\bar{c}(k_T)$, is nearly
constant for the subgraphs of the real topologies, but it grows for
their randomized counterparts. We also experimented with airport networks~\cite{Barrat:2004b} and found that they exhibit
qualitatively the same results as the Internet (BGP) and the social
(PGP) networks. The BGP and PGP networks are more interesting and
challenging for our purposes since, as opposed to airport networks,
they appear to be not explicitly embedded in any observable physical space~\footnote{In the BGP network, nodes are Autonomous Systems that may consist of a large number of routers spreading countries and even continents. Therefore, the physical geography plays only a mild role in the network formation.}.

The high levels of clustering observed in real networks and their self-similarity under the degree thresholding renormalization find a plausible explanation in the
assumption that some metric structures underlay the observed network
topologies. Indeed, under this assumption, clustering becomes
a natural consequence of the triangle inequality in the metric space underneath. The fact that the randomized networks are not self-similar (cf.\
Fig.~\ref{fig2} {\bf b},{\bf d}) also supports this observation.
The applied degree-preserving randomization is a process that involves pairs of nodes, whereas the triangle inequality concerns node triplets.
Therefore, this randomization process cannot fully preserve the
network properties imposed by the triangle inequality.

In the rest of the paper, we provide further evidence that this
metric space explanation is indeed plausible. We do so by
introducing a class of network models designed with the following
three objectives: we want all nodes to exist in a metric space
underlying the network topology; we want to control
the degree distribution and clustering, so that we can generate scale-free graphs with strong clustering; and we want these graphs be small-world.
We then find that the networks generated by our model reproduce all the self-similar effects that we have empirically observed in real networks. We emphasize that although there are models of scale-free networks embedded in Euclidean lattices~\cite{Rozenfeld:2002}, none can simultaneously reproduce all the effects discussed above.

To define our model, we use the hidden variables
formalism~\cite{Boguna:2003b} taking as hidden variables nodes' coordinates in a metric space. Each two nodes are located at a certain hidden metric
distance $d$, and connected with a probability $r$, which relates the network topology to the underlying metric space. This probability depends on the metric
distance $d$ as $r(d/d_c)$, where $d_c$ is the characteristic distance
scale, i.e., a parameter that calibrates whether a given distance is short or long. Function $r$ must be a positive integrable function of $d\in[0,\infty)$.
Consequently, nodes that are close to each other in the metric space
are more likely to be connected in the graph.

To engineer full control over the degree distribution, we link the
characteristic distance scale $d_c$ to the topology of the network.
We assume that $d_c$ is not a constant but depends on some
topological properties of the nodes. Specifically, we assign an
additional hidden random variable $\kappa$ to each node, which corresponds to its expected degree. For simplicity, let our
hidden metric space be a homogeneous and isotropic $D$-dimensional
space. Then the choice~\cite{Serrano:2005a}
\begin{equation}
d_c(\kappa,\kappa') \propto (\kappa \kappa')^{1/D}
\label{eq:1}
\end{equation}
guarantees that the average degree of nodes with variable $\kappa$
is $\bar{k}(\kappa)= \kappa$. Therefore, the distribution
$\rho(\kappa)$ of this variable is asymptotically equal to the degree distribution
$P(k)$ in the resulting networks~\cite{Boguna:2003b}.

Eq.~(\ref{eq:1}) has another important consequence: high degree
nodes --hubs-- are likely to be connected regardless their distance
in the metric space because $d_c(\mbox{hub,hub})$ is large. Low
degree nodes, on the other hand, are connected only if they are
close, whereas hubs are connected to low degree nodes if their
distances are at most intermediate. This pattern is typical of real
networks embedded in metric spaces, such as the airport network. It
is very likely that two major hubs like New York and London are
connected, but it is very unlikely that two small airports are
connected, unless they are close enough.

We can now generate graphs with any degree distribution by choosing
an appropriate $\rho(\kappa)$. In particular, to generate scale-free
graphs we set $\rho(\kappa) =(\gamma-1) \kappa_0^{\gamma-1}
\kappa^{-\gamma}$, $\kappa
> \kappa_0\equiv (\gamma-2) \langle k \rangle/(\gamma-1)$, $\gamma > 2$, which after transformations described in
\cite{Boguna:2003b} yields the degree distribution
\begin{equation}\label{eq:P(k)}
P(k)=(\gamma-1)\kappa_0^{\gamma-1}\frac{\Gamma(k+1-\gamma,\kappa_0)}{k!},
\end{equation}
where $\Gamma$ is the incomplete gamma function. The asymptotic
behavior of this degree distribution for large $k$ is $P(k)\sim
k^{-\gamma}$, i.e., the same as of $\rho(\kappa)$. This result is
independent of the dimension of the hidden metric space and it is
valid for any integrable connection probability of the form
$r(d/d_c)$, where $d_c \equiv d_c(\kappa,\kappa')$ is any
function that factorizes in terms of $\kappa$ and $\kappa'$.

Given the freedom in the choice of the space dimension and particular form of $r$, we hereafter consider the simplest one-dimensional model. We place nodes on a circle,
$\mathbb{S}^1$, by assigning them a random variable $\theta$
representing their polar angle uniformly distributed in the interval
$[0,2\pi)$. The circle radius $R$ grows linearly with the total
number of nodes $N$, $2 \pi R= N/\delta$, in order to keep the
average density of nodes on the circle fixed to a constant value $\delta$ that, without loss of generality, we set to $\delta=1$. We specify the connection probability $r$ such that we can control clustering
in the generated graphs. Specifically, we define $r$, compliant with
Eq.~(\ref{eq:1}), as
\begin{equation}\label{eq:r-final}
r(\theta,\kappa;\theta',\kappa')= \left(
1+\frac{d(\theta,\theta')}{\mu \kappa \kappa'} \right)^{-\alpha},
\quad \alpha>1,
\end{equation}
where $d(\theta,\theta')$ is the geodesic distance over the circle,
i.e., the metric distance $d$ between nodes discussed above,
and $\mu=\frac{(\alpha-1)}{2 \langle k
\rangle}$ is given by the normalization condition $\langle k
\rangle = N/(2\pi)^2 \int \rho(\kappa)
r(\theta,\kappa;\theta',\kappa') \rho(\kappa') d\theta
d\kappa d\theta' d\kappa'$ for large $N$. Parameter $\alpha$
controls clustering. The larger $\alpha$, the more preferred
short-distance connections; thus, the more triangles are formed. The
exact dependence of the average clustering $\bar{c}$ on $\alpha$ is
not important, but both our analytic and simulation results confirm
that, as expected, $\bar{c} \to 0$ when $\alpha \to 1$, and that
$\bar{c}$ converges to a constant value, dependent on $\gamma$, when
$\alpha \to \infty$. We skip the details for brevity; they will be published elsewhere.

We next check whether our synthetic graphs are small worlds, as
spatially embedded networks do not always have this
property~\cite{Gastner:2006,Rozenfeld:2002}. To this end, we compute
the probability $p(d,\kappa|\kappa')$ that a node with hidden
variable $\kappa'$ has a neighbor with hidden variable $\kappa$ at
geodesic distance $d$ on $\mathbb{S}^1$. We use the hidden variables
formalism \cite{Boguna:2003b} and the result reads
\begin{equation}
p(d,\kappa|\kappa')=\frac{2}{\kappa'} \rho(\kappa)
\left(
1+\frac{d}{\mu \kappa \kappa'}
\right)^{-\alpha}.
\end{equation}
Integration over $\kappa$ gives the probability that a node has a
neighbor at distance $d$. For large $d$, this function scales as
$p(d|\kappa') \sim d^{-\alpha}$ when $\alpha<\gamma-1$ and
$p(d|\kappa') \sim d^{1-\gamma}$ when $\alpha>\gamma-1$. The network
is a small world when the average hidden distance to nearest
neighbors $\bar{d}(\kappa') = \int xp(x|\kappa')dx$ diverges in the
large-$N$ and, consequently, large-$R$ limit. Such divergence
indicates the presence of links connecting nodes located at all,
including arbitrarily large, hidden distance scales. This average
distance $\bar{d}(\kappa')$ diverges when the exponent of the
asymptotic form of $p(d|\kappa')$ for large $d$ is smaller than $2$,
i.e., when either $1<\alpha<2$ or $2<\gamma<3$, or both. Real
scale-free networks have values of $\gamma$ between $2$ and $3$,
meaning that they correspond to the class of small-world networks in
our model, regardless of the value of $\alpha$.

\begin{figure}[t]
\begin{center}
\epsfig{file=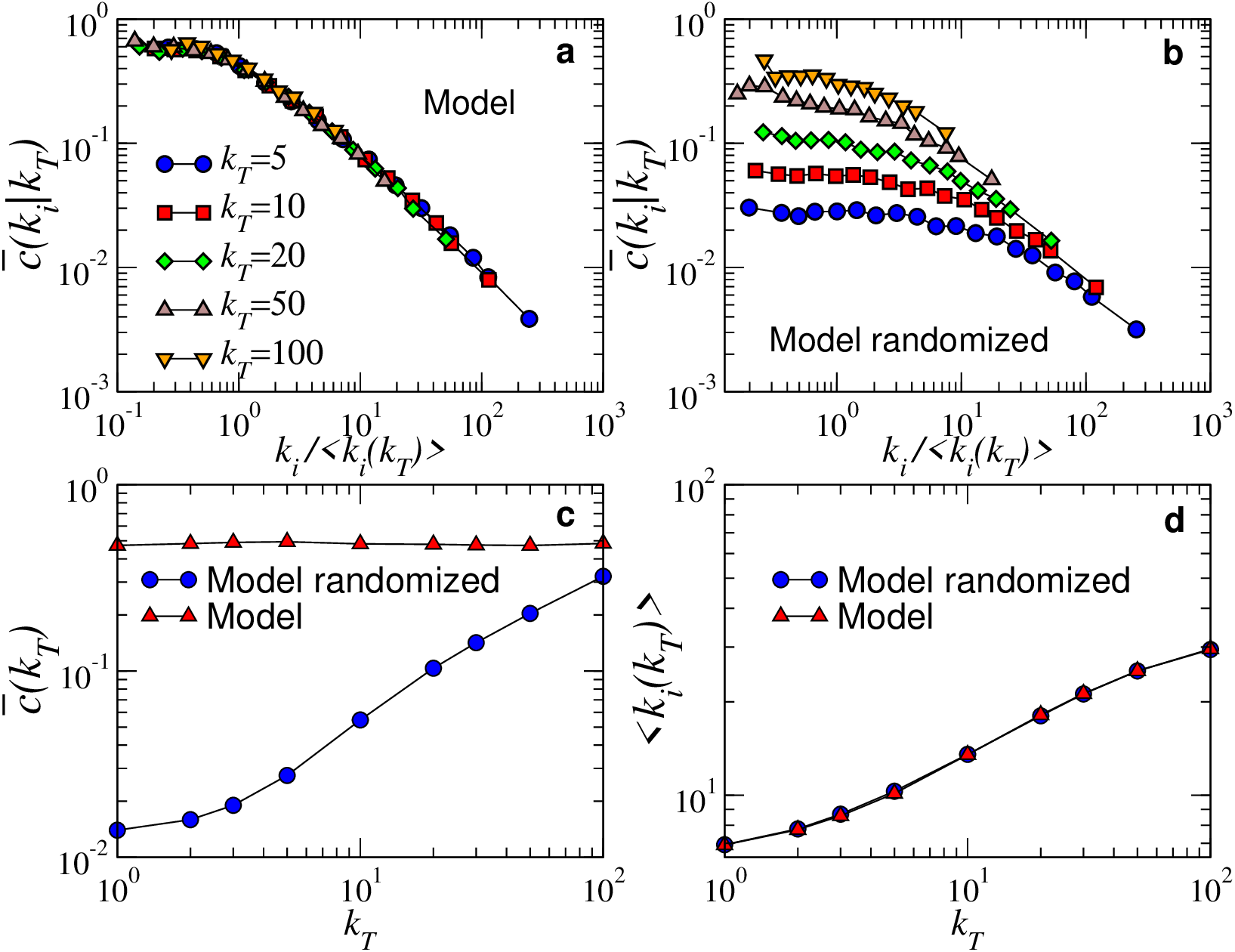,width=8.6cm}
\end{center}
\vspace{-0.5cm} \caption{{\bf a-b:} Scaling of the degree-dependent
clustering coefficient in a modeled network using the connection probability given by Eq. (\ref{eq:r-final}) ($\gamma=2.5$, $\alpha=5.0$, $\langle k \rangle =6$, $N=10^5$)
and its randomization for different values of $k_T$. Average
clustering coefficient, {\bf c}, and average internal degree, {\bf
d}, for the same networks, cf.~Fig.~\ref{fig2}.} \label{fig3}
\vspace{-0.5cm}
\end{figure}

Finally, we analyze the self-similarity of networks produced by our model.
Thanks to the proportionality between $\bar{k}(\kappa)$ and
$\kappa$, we can work in the $\kappa$-space instead of the
$k$-space. Since $\rho(\kappa)$ is a power law,
the distribution of $\kappa$ for nodes in the subgraph $G(\kappa_T)$ (with $\kappa>\kappa_T$),
$\rho(\kappa|\kappa_{T})$, is given by the same power-law function
but starting at $\kappa_T$ instead of $\kappa_0$. Therefore, the average of
$\kappa$ within $G(\kappa_T)$ is given by $\langle \kappa(\kappa_T)
\rangle=\langle k \rangle \kappa_T$. The number of nodes in
$G(\kappa_T)$ is $N\kappa_T^{1-\gamma}$, so that their density in
$\mathbb{S}^1$ is $\delta(\kappa_T)=\delta \kappa_T^{1-\gamma}$,
where $\delta$ is the density of the original graph $G$. Since the
connection probability does not depend on $\kappa_T$, subgraphs
$G(\kappa_T)$ are replicas of $G$ after the following
renormalization of the parameters:
\begin{equation}
\kappa_0 \longrightarrow \kappa_T \hspace{0.5cm}; \hspace{0.5cm}\delta
\longrightarrow \delta \kappa_T^{1-\gamma}.
\end{equation}
In particular, the average degree of nodes with hidden variable
$\kappa$ in $G(\kappa_T)$, $\bar{k}_{i}(\kappa|\kappa_{T})$, and the
average degree of all nodes in $G(\kappa_T)$, $\langle k_{i}
(\kappa_T) \rangle$, are
\begin{equation}
\bar{k}_{i}(\kappa|\kappa_{T})=\kappa_{T}^{2-\gamma}  \kappa
\mbox{\hspace{0.3cm}and\hspace{0.3cm}} \langle k_{i} (\kappa_T)
\rangle =\kappa_T^{3-\gamma} \langle k \rangle. \label{k_internal}
\end{equation}
The degree distribution in $G(\kappa_T)$ is given by the same
analytic expression Eq.~(\ref{eq:P(k)}), except that $\langle k_{i}
(\kappa_T) \rangle$ from Eq.~(\ref{k_internal}) replaces $\langle k
\rangle$ in Eq.~(\ref{eq:P(k)}).

The exact expression for clustering in $G(\kappa_T)$ is rather long
and we omit it here for brevity, but it can be easily derived from results in~\cite{Boguna:2003b}. What matters
for our analysis is that the clustering coefficient of nodes with
hidden variable $\kappa$ in $G(\kappa_T)$ satisfies
$\bar{c}(\kappa|\kappa_{T})=f(\kappa/\kappa_{T})$, where $f$ is some
function. It follows that the average clustering coefficient in
$G(\kappa_T)$, $\bar{c}(\kappa_T)=\int_{\kappa_{T}} d\kappa
\rho(\kappa|\kappa_{T})\bar{c}(\kappa|\kappa_{T})$, takes a finite
value independent of $\kappa_{T}$. Using once again the
proportionality between the $\kappa$-space and $k$-space and the
scaling relations in Eq.~(\ref{k_internal}), we conclude that
\begin{equation}
\bar{c}(k_{i}|k_{T})\approx
f(k_{i}/k_{T}^{3-\gamma})=\tilde{f}(k_{i}/\langle k_{i} (k_T)
\rangle), \label{scaling_clustering}
\end{equation}
where we use the symbol ``$\approx$'' to account for fluctuations of
degrees of nodes with small values of $\kappa$.

In Fig.~\ref{fig3}, we show that our simulation results match
perfectly the scaling of clustering predicted by
Eq.~(\ref{scaling_clustering}). The same figure demonstrates that
the clustering-related self-similarity properties of our modeled
networks and their randomizations are qualitatively the same as of
real networks in Fig.~\ref{fig2}. We emphasize that the self-similarity of clustering observed in our model does not depend either on the
dimension of the hidden space or on the final form of $r$. The only
requirements are that nodes are located in a metric space and
connected under the integrable connection probability $r(d/d_c)$
with $d_c$ given by Eq.~(\ref{eq:1}), and that the degree distribution is scale-free.

In summary, hidden geometries underlying the
observed topologies of some complex networks appear to provide a
simple and natural explanation of their degree-renormalization
self-similarity. If we
take the most generic interpretation of hidden distances as measures
of either structural or functional similarity between
nodes~\cite{WatDoNew02,menczer02-pnas}, and admit that more similar
nodes are more likely to be connected, then the hidden and
observable forms of transitivity become clearly related. At the hidden geometry layer, this transitivity is the transitivity of ``being close,'' while at the observed
topology layer, it is the transitivity of ``being connected.'' In future work, hidden metric spaces may find far-reaching applications such as the design of efficient
routing and searching algorithms for communication and social
networks. Also worth pursuing is studying
the relationship between fractality in~\cite{Makse:2005}
and self-similarity under our
renormalization procedure.

\begin{acknowledgments}
We thank Romualdo Pastor-Satorras and kc claffy for useful suggestions.
This work was supported by DELIS FET Open 001907 and the SER-Bern
02.0234, by NSF CNS-0434996 and CNS-0722070, and by DGES FIS2004-05923-CO2-02, FIS2007-66485-C02-01, and FIS2007-66485-C02-02, Generalitat de
Catalunya grant No. SGR00889.
\end{acknowledgments}
\vspace{-0.5cm}

%\bibliographystyle{apsrev}
%\bibliography{ref,bib}

\end{document}